\documentclass[12pt]{article}
\usepackage{epsfig}

\begin{document}

\begin{center}

\vspace*{1.0cm}

 {\Large \bf{First search for double-$\beta$ decay of $^{184}$Os and $^{192}$Os}}

\vskip 1.0cm

{\bf P.~Belli$^{a}$, R.~Bernabei$^{a,b,}$\footnote{Corresponding
author. {\it E-mail address:} rita.bernabei@roma2.infn.it
(R.~Bernabei).}, F.~Cappella$^{c,d}$, R.~Cerulli$^{e}$,
F.A.~Danevich$^{f}$, S.~d'Angelo$^{a,b}$, A.~Di~Marco$^{a,b}$,
A.~Incicchitti$^{c,d}$, G.P.~Kovtun$^{g}$, N.G.~Kovtun$^{g}$,
M.~Laubenstein$^{e}$, D.V.~Poda$^{f}$, O.G.~Polischuk$^{c,f}$,
A.P.~Shcherban$^{g}$, V.I.~Tretyak$^{f}$}

\vskip 0.3cm

$^{a}${\it INFN, sezione Roma ``Tor Vergata'', I-00133 Rome,
Italy}

$^{b}${\it Dipartimento di Fisica, Universit$\grave{a}$ di Roma
``Tor Vergata'', I-00133 Rome, Italy}

$^{c}${\it INFN, sezione Roma ``La Sapienza'', I-00185 Rome, Italy}

$^{d}${\it Dipartimento di Fisica, Universit$\grave{a}$ di Roma ``La Sapienza'', I-00185 Rome,
Italy}

$^{e}${\it INFN, Laboratori Nazionali del Gran Sasso, 67100 Assergi (AQ), Italy}

$^{f}${\it Institute for Nuclear Research, MSP 03680 Kyiv, Ukraine}

$^{g}${\it National Science Center ``Kharkiv Institute of Physics
and Technology'', 61108 Kharkiv, Ukraine}

\end{center}

\vskip 0.5cm

\begin{abstract}
A search for double-$\beta$ decay of osmium has been realized for the
first time with the help of an ultra-low background HPGe $\gamma$
detector at the underground Gran Sasso National Laboratories of
the INFN (Italy). After 2741 h of data taking with a 173 g
ultra-pure osmium sample limits on double-$\beta$ processes in
$^{184}$Os have been established at the level of $T_{1/2}\sim 10^{14}-10^{17}$ yr.
Possible resonant double-electron captures in $^{184}$Os were searched
for with a sensitivity $T_{1/2}\sim 10^{16}$ yr. A half-life limit
$T_{1/2}\geq 5.3\times 10^{19}$ yr was set for the double-$\beta$ decay
of $^{192}$Os to the first excited level of $^{192}$Pt. The radiopurity
of the osmium sample has been investigated and radionuclides $^{137}$Cs,
$^{185}$Os and $^{207}$Bi were detected in the sample, while
activities of $^{40}$K, $^{60}$Co, $^{226}$Ra and $^{232}$Th were limited at the
$\approx$ mBq/kg level.

\end{abstract}

\vskip 0.4cm


\vskip 0.4cm


\section{INTRODUCTION}

Neutrinoless double-$\beta$ decay ($0\nu2\beta$) is one of the most promising probes for physics beyond the Standard Model. The process is 
sensitive to lepton number violation,
the nature of the neutrino (Majorana or Dirac particle), the absolute
neutrino mass and the neutrino mass hierarchy (see
\cite{Zdes02,Elli02,Verg02,Elli04,Avig05,Ejir05,Avig08,Klap08,Rode11,Elli12,Verg12}
and references therein). In particular, the investigation of
neutrinoless double-electron capture ($0\nu2\varepsilon$) and of
 electron capture with positron emission
($0\nu\varepsilon\beta^{+}$) could clarify the possible
contribution of the right-handed currents to the $0\nu2\beta^{-}$
decay rate \cite{Klap08,Hirs94}.

The Osmium contains two potentially double-$\beta$ active isotopes:
$^{184}$Os (energy of decay $Q_{2\beta}=1453.7(0.6)$ keV
\cite{Smor12}; isotopic abundance $\delta= 0.02(1)$\%
\cite{Berg11}; allowed decay channels: $2\varepsilon$ and
$\varepsilon\beta^{+}$) and $^{192}$Os ($Q_{2\beta}=412.4(2.9)$
keV \cite{Audi03}; $\delta=40.78(19)$\% \cite{Berg11};
$2\beta^{-}$). The decay schemes of $^{184}$Os and $^{192}$Os are
presented in Figs.~1 and 2, respectively. There is a possibility
of a resonant enhancement of the $0\nu$ double-electron capture in
$^{184}$Os to a few excited levels of $^{184}$W. The most
promising of them is the level $(0)^{+}$ 1322.2 keV \cite{Kriv11}.

\begin{figure}[htb]
\begin{center}
 \mbox{\epsfig{figure=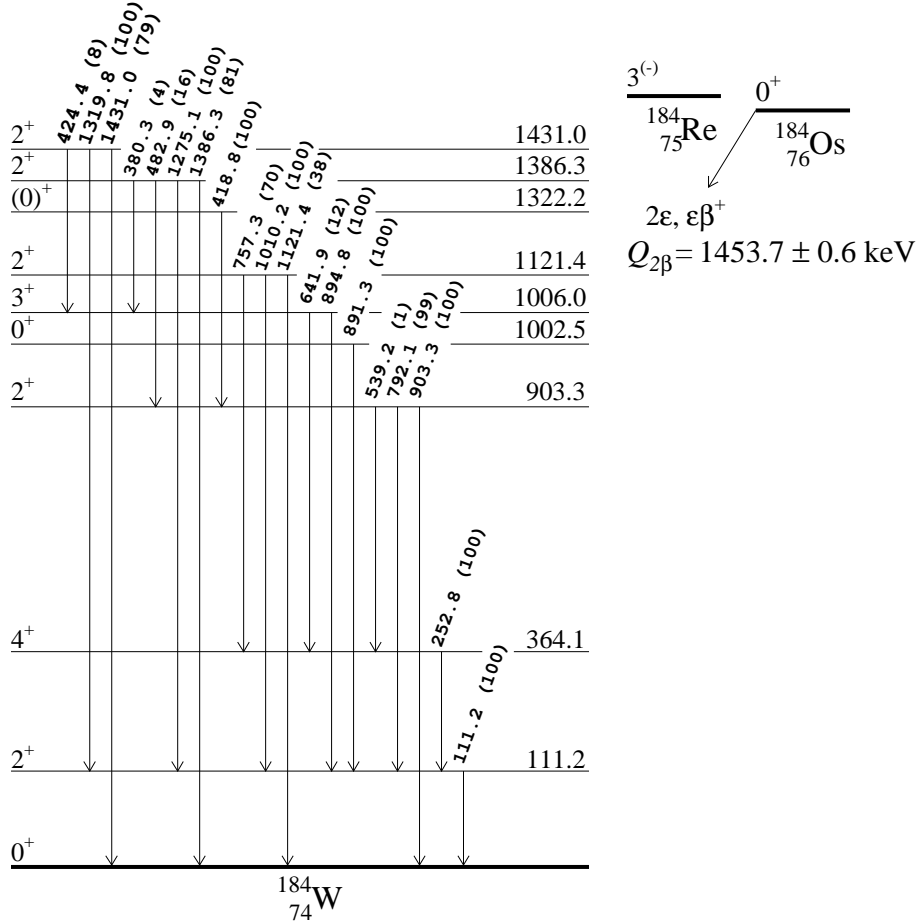,height=12.5cm}}
\vspace{-0.3cm}
\caption{Decay scheme of $^{184}$Os \cite{Bagl10a}. The energies
of the excited levels and of the emitted $\gamma$ quanta are in keV
(relative intensities of $\gamma$ quanta, rounded to percent, are
given in parentheses).}
\end{center}
\end{figure}
\vspace{-0.1cm}

\begin{figure}[htb]
\begin{center}
 \mbox{\epsfig{figure=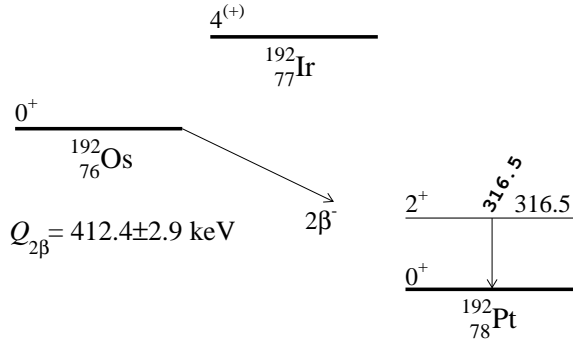,height=4.7cm}}
\vspace{-0.3cm}
\caption{Decay scheme of $^{192}$Os \cite{Bagl10b}.}
\end{center}
\end{figure}
\vspace{-0.3cm}

We have investigated the radiopurity of the osmium sample and realized the
first search for double-$\beta$ processes in $^{184}$Os and $^{192}$Os
with the help of an ultra-low background
HPGe $\gamma$ spectrometer. Preliminary results of the experiment
were presented in \cite{Bell12}.

\section{MEASUREMENTS, RESULTS AND DISCUSSION}

\subsection{Experiment}

An ultra-pure osmium sample (more than 99.999\% purity grade
\cite{Azha00}) with a mass of 173 g was used in the experiment. The
material was obtained using electron-beam melting of osmium
powder with further purification by electron-beam zone
refining. The estimated density of the metal is 23 g/cm$^3$ (the
tabulated value is 22.57 g/cm$^3$ \cite{Lide03}). The sample contains
$1.1\times10^{20}$ nuclei of $^{184}$Os and $2.2\times10^{23}$
nuclei of $^{192}$Os assume the natural isotopic composition \cite{Berg11}. The experiment was carried out at the Gran
Sasso National Laboratories of the INFN (Italy) with the ultra-low
background HPGe detector GeCris with a volume of 465 cm$^3$. The
detector is shielded by lead ($\approx 25$ cm) and copper ($\approx
10$ cm). The FWHM energy resolution
of the spectrometer is 2.0 keV for the 1333 keV $\gamma$
quanta of $^{60}$Co. The data with the sample were accumulated
over 2741 h, while the background spectrum was taken over 1046 h.
The spectra are presented in Fig. 3.

\begin{figure}[htb]
\begin{center}
 \mbox{\epsfig{figure=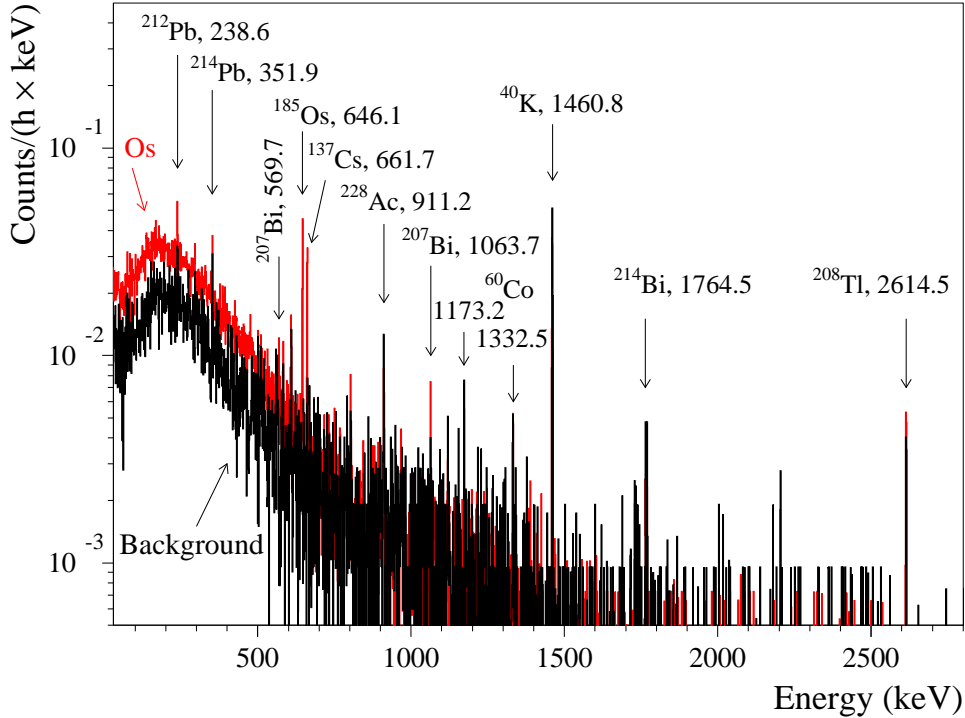,height=10.0cm}}
\vspace{-0.3cm}
\caption{(Color online) Energy spectra measured with the ultra-low background HPGe $\gamma$ spectrometer
with the osmium sample over 2741 h (Os) and without sample over 1046 h
(Background). The energies of the $\gamma$ lines are in keV.}
\end{center}
\end{figure}
\vspace{-0.3cm}

\subsection{Radioactive contamination of the osmium sample}

The peaks in the energy distributions can be ascribed to $\gamma$
quanta of U/Th daughters, $^{40}$K, $^{60}$Co,
$^{137}$Cs, $^{185}$Os and $^{207}$Bi. The response
functions of the detector to decays of these nuclides as well as
to the double-$\beta$ processes in the osmium isotopes were
simulated by EGSnrc \cite{EGSnrc} and GEANT4 \cite{Agos03,Alli06}
packages with initial kinematics given by the DECAY0 event
generator \cite{DECAY0}. Both simulations gave consistent
results\footnote{For example, the detection efficiency to the
$\gamma$ quanta with the energy 418.8 keV, expected to be emitted
in the resonant $0\nu2\varepsilon$ process in $^{184}$Os, is
3.14\% (3.12\%) with EGSnrc (GEANT4).}. Some excess of events
in the 662 keV peak of $^{137}$Cs was observed with an activity
($1.9\pm0.3$) mBq/kg. The radioactive $^{185}$Os (electron capture
decay, $T_{1/2}=93.6$ d \cite{ToI98}) was also detected in the
sample with an activity ($3.0\pm0.3$) mBq/kg. We assume that this
radionuclide was generated before the installation of the sample
into the set-up by capture of neutrons by $^{184}$Os and by
spallation processes induced by cosmic rays on heavier osmium
isotopes. The presence of $^{207}$Bi at the level of ($0.4\pm0.1$)
mBq/kg can be explained by contamination of the sample.
The activities of $^{137}$Cs, $^{185}$Os and $^{207}$Bi, as well
as upper limits of $^{40}$K, $^{60}$Co and U/Th daughters are
presented in Table 1.

\nopagebreak
\begin{table}[!h]
\caption{Radioactive impurities in the osmium sample. The upper
limits are presented at 90\% CL, the uncertainties are given with
68\% CL. The reference date for activities is October 2011.}
\begin{center}
\begin{tabular}{|l|l|l|}

 \hline
  Chain     & Nuclide       & Activity (mBq/kg) \\
  \hline
 $^{232}$Th & $^{228}$Ra    & $\leq 2.0$ \\
 ~          & $^{228}$Th    & $\leq 2.3$ \\
 \hline
 $^{238}$U  & $^{226}$Ra    & $\leq 0.6$  \\
\hline
 ~          & $^{40}$K      & $\leq 1.9$  \\
 ~          & $^{60}$Co     & $\leq 0.1$  \\
 ~          & $^{137}$Cs    & $1.9\pm0.3$ \\
 ~          & $^{185}$Os    & $3.0\pm0.3$ \\
 ~          & $^{207}$Bi    & $0.4\pm0.1$ \\

\hline
\end{tabular}
\end{center}
\end{table}

\subsection{Search for $\varepsilon\beta^{+}$ and $2\varepsilon$ processes in $^{184}$Os}

We do not observe any peaks in the energy distribution accumulated with the
osmium sample which could indicate double-$\beta$ activity of
$^{184}$Os or $^{192}$Os. Therefore only lower half-life limits
($\lim T_{1/2}$) can be set according to the formula:

\begin{center}
$$\lim T_{1/2} = N \cdot \eta \cdot t \cdot \ln 2 / \lim S,$$
\end{center}
where $N$ is the number of potentially $2\beta$ unstable nuclei,
$\eta$ is the detection efficiency (in which the yields of the specific
$\gamma$ quanta in accordance with the scheme of the decay, Fig. 1, are included),
$t$ is the measuring time, and
$\lim S$ is the upper limit on the number of events of the effect searched for which
can be excluded at a given confidence level (CL). All the
half-life limits and the values of $\lim S$ are presented in this paper at 90\% CL.

One positron can be emitted in the $\varepsilon\beta^{+}$ decay of
$^{184}$Os with an energy up to ($431.7 \pm 0.6$) keV. The
annihilation of the positron should give two 511 keV  $\gamma$'s
leading to an extra rate in the annihilation peak. To estimate
$\lim S$ for the decay, the energy spectra were fitted in the
energy interval ($495-530$)~keV (see Fig.~4). There are ($92\pm24$)
events in the 511 keV peak in the data accumulated with the osmium
sample.
\begin{figure}[htb]
\begin{center}
 \mbox{\epsfig{figure=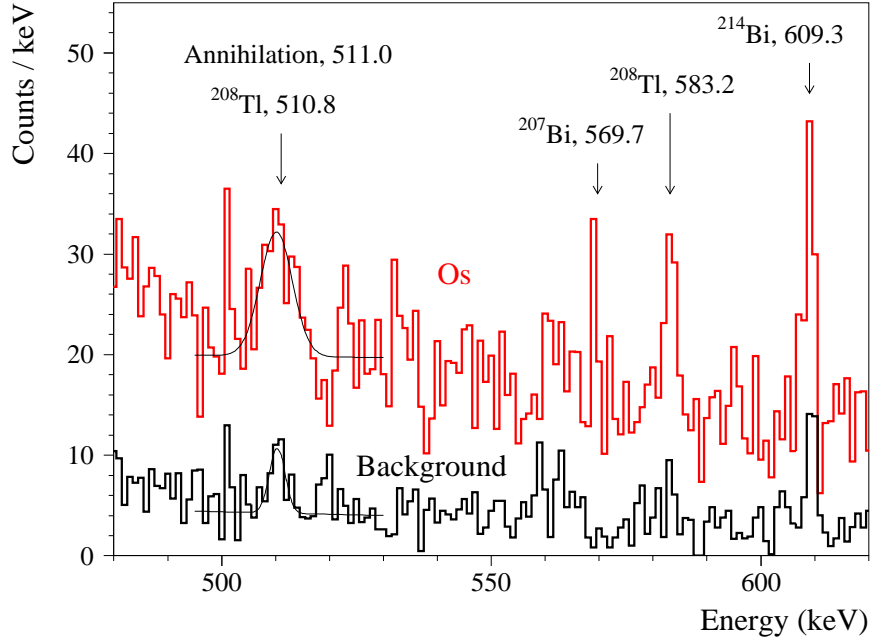,height=8.5cm}}
\vspace{-0.3cm}
\caption{(Color online) Part of the energy spectra accumulated
with the osmium sample over 2741 h (Os sample) and without the sample
over 1046 h (Background) in the vicinity of the annihilation peak.
Fits of the peaks are shown by the solid lines.}
\end{center}
\end{figure}
\vspace{-0.3cm}
The area of the annihilation peak in the background spectrum is
($24\pm10$) counts, corresponding to ($63\pm26$) counts after normalising to the exposure of the osmium sample.
The difference in the areas of
the annihilation peak with and without sample [($29\pm35$) counts]
gives no indication of the effect. In accordance with the
Feldman-Cousins procedure \cite{Feld98} we should take $\lim S=86$
counts which can be excluded at 90\% CL. Taking into account
almost the same detection efficiency for 2$\nu$ ($\eta$=9.0\%) and 0$\nu$
(8.9\%)
modes, we have obtained the following limit on the
half-life of $^{184}$Os for the $2\nu$ and $0\nu$ modes of
$\varepsilon\beta^{+}$ decay (see also Table 2 where the energies
of $\gamma$ quanta used in the analysis, the detection efficiencies
and the values of $\lim S$ are given):

\begin{center}
 $T_{1/2}^{(2\nu+0\nu)\varepsilon\beta^{+}}(^{184}$Os, g.s. $\to$ g.s.)$~\geq 2.5\times 10^{16}$ yr.
\end{center}

The $\varepsilon\beta^+$ decay of $^{184}$Os is also allowed to
the first excited level of $^{184}$W with an energy of 111.2 keV.
The strongest restrictions on the decay were obtained by analysis
of the annihilation peak. Taking into account slightly different
detection efficiencies for $2\nu$ (9.0\%) and $0\nu$ (8.8\%)
modes, we have obtained the following limits:

\begin{center}
 $T_{1/2}^{2\nu\varepsilon \beta^+}(^{184}$Os, g.s.$~\to~$111.2 keV$)\geq 2.5\times10^{16}$ yr,
\end{center}

\begin{center}
 $T_{1/2}^{0\nu\varepsilon \beta^+}(^{184}$Os, g.s.$~\to~$111.2 keV$)\geq 2.4\times10^{16}$ yr.
\end{center}

In the case of the $2\nu2K$ capture in $^{184}$Os, a cascade of X
rays and Auger electrons with individual energies up to 69.5
keV is expected. The most intensive X ray lines of tungsten are
\cite{ToI98}: 58.0 keV (the yield of the X ray quanta is 27.4\%),
59.3 keV (47.0\%), 67.0 keV (5.4\%), 67.2 keV (10.3\%) and 69.1
keV (3.6\%). To derive a limit on the decay, the energy spectrum
was fitted in the energy interval ($47-72$) keV by the model
consisting of five Gaussians (to describe the expected X ray
peaks) plus a polynomial function of the first degree
(background). The fit provides the total area of the
$2\nu2K$ effect ($21\pm26$) counts, which corresponds to $\lim
S=64$ counts (see Fig.~5). The detection efficiency of the whole effect in this
case is calculated as: $\eta = \Sigma_i \ \eta_i$, where $\eta_i$
are the efficiencies for the specific X lines with energies from
58.0 keV to 69.1 keV. Taking into account $\eta =
0.05\%$\footnote{We have calculated the efficiency of the set-up to
detect the X-rays by using both the EGSnrc and GEANT4
codes, which again give consistent results.}, one can
calculate the following half-life limit:

 \begin{center}
 $T_{1/2}^{2\nu2K}(^{184}$Os, g.s.$~\to~$g.s.$)\geq 1.9\times10^{14}$ yr.
 \end{center}

\begin{figure}[htb]
\begin{center}
 \mbox{\epsfig{figure=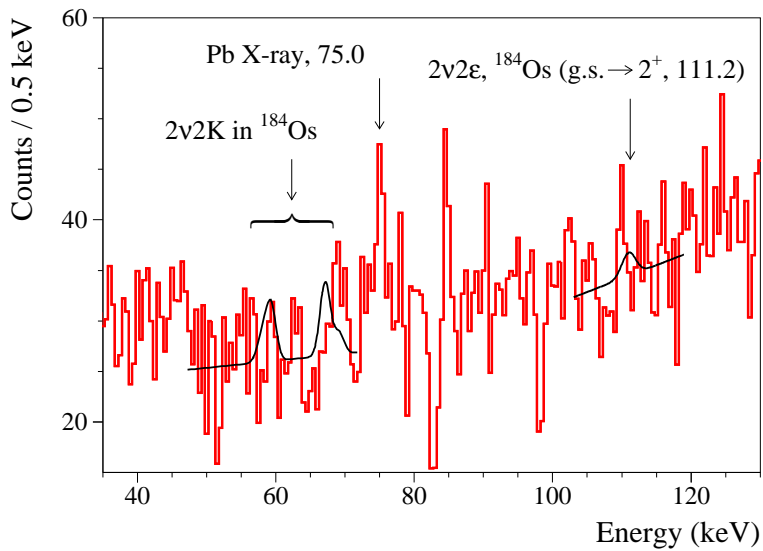,height=7.5cm}}
\vspace{-0.3cm}
\caption{(Color online) Low energy part of the energy spectrum
accumulated with the osmium sample over 2741 h. Excluded effects
of $2\nu2K$ process in $^{184}$Os with the half-life
$T_{1/2}=1.9\times10^{14}$ yr, and of $2\nu2K$ transition to the excited $2^{+}$ level of $^{184}$W with an energy of 111.2 keV with the half-life
$T_{1/2}=3.1\times10^{15}$ yr are shown by the solid lines.}
\end{center}
\end{figure}
\vspace{-0.3cm}

The sensitivity of the experiment to the $2\nu2\varepsilon$ decay
of $^{184}$Os to the 111.2 keV excited level of $^{184}$W is
estimated from the fit of the energy spectrum accumulated with the
osmium sample in the vicinity of the energy 111.2 keV. The fit
gives an area for the $\gamma$ peak of ($-4.8\pm6.1$) counts, that
leads to $\lim S=5.8$ counts (see Fig.~5). The detection efficiency for the
111.2 keV $\gamma$ quanta, taking into account the fact that the
transition from the 111.2 keV excited level to the ground state of
$^{184}$W is efficiently converted to electrons (the experimental
coefficient of conversion is $\alpha_{total}=2.57$
\cite{Bagl10a}), was calculated to be 0.076\%. Generally speaking,
due to the large coefficient of conversion to electrons, the
sensitivity of the experiment should be different for $2K$, $KL$
and $2L$ captures. For example, since both $K$ electrons are
already captured in the $2K$ process, the conversion from the $K$
shell is prohibited in further 111.2 keV deexcitation process. The
coefficients of conversion from different atomic shells can be
calculated with the BrIcc program \cite{Kibe08}\footnote{For the total conversion coefficient it gives a value of 2.57,
in good agreement with the experimental result \cite{Bagl10a}.}. However, the excited
111.2 keV level of $^{184}$W has quite a long half-life of 1.251
ns \cite{Bagl10a} compared to the characteristic atomic relaxation time of
 $\sim 10^{-15}$ s (see e.g.
\cite{Dre02}). Thus, we consider that the relaxation processes in
the W atomic shell after the initial $2\nu2\varepsilon$ capture
are finished before the 111.2 keV deexcitation and we can apply
the value $\alpha = 2.57$ for all $2\nu2\varepsilon$ captures.
This gives:

\begin{center}
 $T_{1/2}^{2\nu2\varepsilon}(^{184}$Os, g.s.$~\to~$111.2 keV$)\geq 3.1\times10^{15}$ yr.
\end{center}

To estimate limits on the $2\nu2\varepsilon$ decay of $^{184}$Os
to the $2^{+}$ 903.3 keV, $0^{+}$ 1002.5 keV and $2^{+}$ 1121.4
keV excited levels of $^{184}$W, the energy spectrum accumulated
with the osmium sample was fitted in the energy intervals where
intense $\gamma$ peaks from the de-excitation process are
expected. For instance, we have obtained the following limit on
the $2\nu2\varepsilon$ decay of $^{184}$Os to the excited 903.3
keV level of $^{184}$W by the fit of the data in the energy interval
($890-910$) keV where the 903.3 keV $\gamma$ peak is expected:

 \begin{center}
 $T_{1/2}^{2\nu2\varepsilon}(^{184}$Os, g.s.$~\rightarrow~903.3~$keV$)\geq 3.2\times10^{16}$ yr.
 \end{center}

The limits obtained for the $2\nu2\varepsilon$ decay of $^{184}$Os
to the excited levels of $^{184}$W are presented in Table~2.

In the neutrinoless double-electron capture in addition to X rays
we suppose that only one bremsstrahlung $\gamma$ quantum is
emitted \cite{Wint55} to take away the rest of the energy, which in the $2\nu$
process is taken by the neutrinos. In the case of the $0\nu$ double-
electron capture from $K$ and $L$ shells, the energy of the
$\gamma$ quanta is expected to be equal to
$E_{\gamma}=Q_{2\beta}-E_{b1}-E_{b2}$, where $E_{bi}$ are the
binding energies of the captured electrons on the $K$ and $L$
atomic shells of tungsten \cite{ToI98}: $E_{K}=69.5$ keV, $E_{L_{1}}=12.1$ keV,
$E_{L_{2}}=11.5$ keV, $E_{L_{3}}=10.2$ keV. Therefore, the
expected energies of the quanta for the $0\nu2\varepsilon$ capture
in $^{184}$Os to the ground state of $^{184}$W are:
$E_{\gamma}=(1314.7\pm0.6)$ keV for $2K$, $E_{\gamma}=(1373.1\pm1.6)$
keV for $KL$ and $E_{\gamma}=(1431.4\pm2.5$) keV for $2L$. There are
no clear peaks in the data with these energies (see Fig.~6). We
have estimated the values of $\lim S$ by fitting the data in the energy
intervals with the simple model: a Gaussian function (to describe
the peaks searched for) plus a polynomial function (continuous
background). Taking into account the calculated efficiencies to
detect $\gamma$ quanta with energies of $\approx(1.31-1.43)$ MeV
($\approx4.1\%~-~4.2\%$), we set the following limits on the
processes searched for:

 \begin{center}
 $T_{1/2}^{0\nu 2K}(^{184}$Os, g.s.$~\rightarrow~$g.s.$)\geq~2.0\times10^{17}$ yr,

 $T_{1/2}^{0\nu KL}(^{184}$Os, g.s.$~\rightarrow~$g.s.$)\geq~1.3\times10^{17}$ yr,

 $T_{1/2}^{0\nu 2L}(^{184}$Os, g.s.$~\rightarrow~$g.s.$)\geq~1.4\times10^{17}$ yr.
 \end{center}

\begin{figure}[htb]
\begin{center}
 \mbox{\epsfig{figure=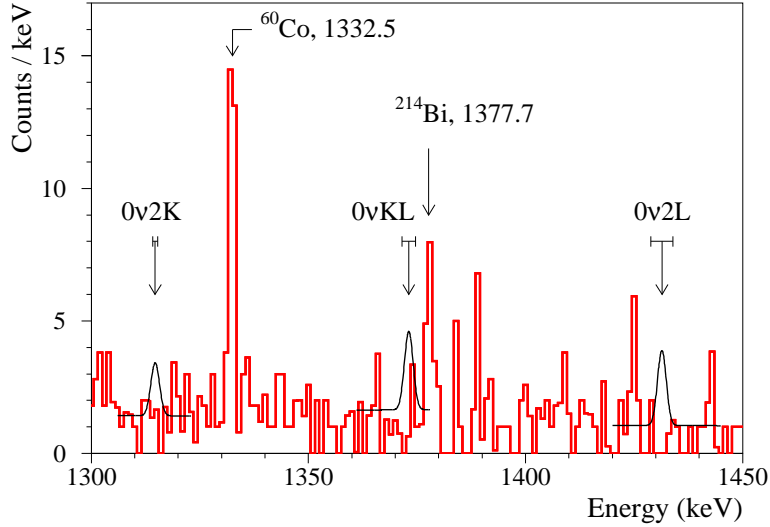,height=7.0cm}}
\vspace{-0.3cm}
\caption{(Color online) Part of the energy spectrum where the
peaks from the $0\nu2\varepsilon$ captures ($2K$, $KL$ and $2L$)
in $^{184}$Os to the ground state of $^{184}$W are expected. The peaks that are excluded at 90\% CL
 are shown by the solid lines.}
\end{center}
\end{figure}
\vspace{-0.3cm}

The $Q_{2\beta}$ energy of $^{184}$Os allows also the population
of several excited levels of $^{184}$W. The limits obtained for
the processes of $0\nu2\varepsilon$ decay to the ground and
to the excited levels of the daughter nuclei are presented in Table 2.

\subsection{Resonant double-electron capture in $^{184}$Os}

The neutrinoless double-electron capture of $^{184}$Os to the excited levels of
$^{184}$W with energies of 1322.2 keV, 1386.3 keV and 1431.0 keV
may occur with higher probability due to a possible resonant
enhancement. The double capture on the $(0)^{+}$ 1322.2 keV level
was considered as the most promising with a half-life in the
range ($7\times10^{26}-2\times10^{27}$) yr for the Majorana
neutrino mass of 1 eV \cite{Kriv11}\footnote{However, following a recent
 high-precision measurement of the $Q$ value for the double-
$\beta$ decay of $^{184}$Os
\cite{Smor12}, the authors have concluded
that the half-life of the transition exceeds $1.3\times10^{29}$
yr.}. Limits on resonant double-electron capture in
$^{184}$Os from $K$ and $L$ shells were obtained by analyzing the
experimental spectrum in the energy intervals where the most
strongest $\gamma$ lines from the de-excitation of these levels are
expected. For example, the greatest sensitivity to the resonant
$2K$ capture in $^{184}$Os to the excited $(0)^+$ level of
$^{184}$W with an energy of 1322.2 keV is obtained by searching for the de-excitation $\gamma$-ray
with an energy 903.3 keV, yielding the limit:

\begin{center}
 $T_{1/2}^{Res~0\nu2K}(^{184}$Os, g.s.$~\rightarrow~1322.2$ keV$)\geq 2.8 \times 10^{16}$ yr.
\end{center}

Limits on the resonant double-electron captures to the excited levels
$2^{+}$ 1386.3 keV and $2^{+}$ 1431.0 keV were obtained in a
similar way. The half-life limits on the resonant $2\varepsilon$
processes in $^{184}$Os are presented in Table~2.

\subsection{Double-$\beta^{-}$ decay of $^{192}$Os}

To set a limit on the $2\beta^{-}$ transition of $^{192}$Os to the
$2^{+}$ 316.5 keV excited level of $^{192}$Pt, the energy spectrum
was fitted by a straight line (which represents the background
model) and a Gaussian at 316.5 keV. The
energy spectrum in the vicinity of the peak is presented in
Fig.~7. Taking into account the detection efficiency ($\eta=3.0\%$
both for the $2\nu$ and $0\nu$ modes of the decay), we have
obtained the following limit:

\begin{center}
$T_{1/2}^{(2\nu+0\nu)2\beta^{-}}(^{192}$Os,
g.s.$~\rightarrow~316.5$ keV$)\geq~5.3\times10^{19}$ yr.
\end{center}

\begin{figure}[t]
\begin{center}
 \mbox{\epsfig{figure=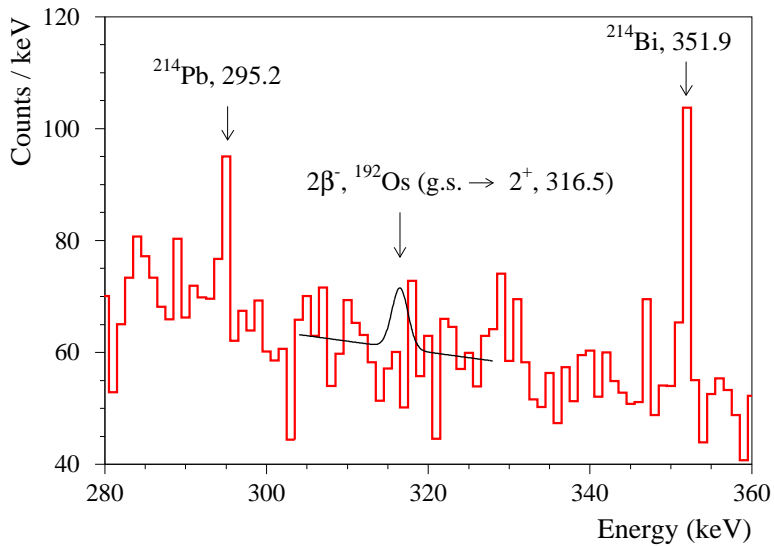,height=7.5cm}}
\vspace{-0.3cm}
\caption{(Color online) Part of the energy spectrum where the peak
from the $2\beta$ decay of $^{192}$Os to the first excited level of
$^{192}$Pt is expected. The excluded -- at 90\% CL -- peak is shown by
the solid line.}
\end{center}
\end{figure}
\vspace{-0.3cm}

\begin{table*}[htbp]
\vspace{-0.7cm}
\caption{Half-life limits on 2$\beta$ processes in $^{184}$Os and
$^{192}$Os.}
\begin{center}
\resizebox{1.00\textwidth}{!}{
\begin{tabular}{|l|l|l|l|l|l|l|}
\hline
 Process                    & Decay     & Level of      & $E_\gamma$  & Detection   & $\lim S$  & Experimental \\
 of decay                   & mode      & daughter      & (keV)       & efficiency  & ~         & limit (yr) \\
 ~                          & ~         & nucleus       &             & ~           &  ~        & at 90\% CL \\
 ~                          & ~         & (keV)         &             & ~           &  ~        &  ~\\
 \hline
 $^{184}$Os$~\to$$^{184}$W   & ~         & ~             & ~           & ~          &  ~        &  ~\\
 $\varepsilon\beta^+$       & $2\nu$    & g.s.          & 511         & 9.0\%       & 86        & $\geq2.5\times10^{16}$ \\
 $\varepsilon\beta^+$       & $0\nu$    & g.s.          & 511         & 8.9\%       & 86        & $\geq2.5\times10^{16}$ \\
 $\varepsilon\beta^+$       & $2\nu$    & $2^{+}~111.2$ & 511         & 9.0\%       & 86        & $\geq2.5\times10^{16}$ \\
 $\varepsilon\beta^+$       & $0\nu$    & $2^{+}~111.2$ & 511         & 8.8\%       & 86        & $\geq2.4\times10^{16}$ \\

 $2K$                       & $2\nu$    & g.s.          & $58-69$     & 0.05\%      & 64        & $\geq1.9\times10^{14}$ \\
 $2\varepsilon$             & $2\nu$    & $2^{+}~111.2$ & 111.2       & 0.076\%     & 5.8       & $\geq3.1\times10^{15}$ \\
 $2\varepsilon$             & $2\nu$    & $2^{+}~903.3$ & 903.3       & 2.3\%       & 17        & $\geq3.2\times10^{16}$ \\
 $2\varepsilon$             & $2\nu$    & $0^{+}~1002.5$& 891.3       & 4.5\%       & 2.8       & $\geq3.8\times10^{17}$ \\
 $2\varepsilon$             & $2\nu$    & $2^{+}~1121.4$& 757.3       & 1.5\%       & 5.2       & $\geq6.9\times10^{16}$ \\

 $2K$                       & $0\nu$      & g.s.          & $1314.1-1315.3$ & 4.1\% & 5.0        & $\geq2.0\times10^{17}$ \\
 $KL$                       & $0\nu$      & g.s.          & $1371.5-1374.7$ & 4.1\% & 7.4        & $\geq1.3\times10^{17}$ \\
 $2L$                       & $0\nu$      & g.s.          & $1428.9-1433.9$ & 4.2\% & 7.1        & $\geq1.4\times10^{17}$ \\

 $2K$                       & $0\nu$      & $2^+$  111.2  & $1202.9-1204.1$ & 4.2\%       & 2.7        & $\geq3.3\times10^{17}$ \\
 $2\varepsilon$             & $0\nu$      & $2^+$  903.3   & 903.3    & 2.0\%       & 17         & $\geq2.8\times10^{16}$ \\
 $2\varepsilon$             & $0\nu$      & $0^+$  1002.5  & 891.3    & 4.1\%       & 2.8        & $\geq3.5\times10^{17}$ \\
 $2\varepsilon$             & $0\nu$      & $2^+$ 1121.4   & 757.3    & 1.4\%       & 5.2        & $\geq6.4\times10^{16}$ \\

 Resonant $2K$             & $0\nu$     & $(0)^+$ 1322.2  & 903.3    & 2.0\%       & 17          & $\geq2.8\times10^{16}$ \\
 Resonant $KL$             & $0\nu$     & $2^+$ 1386.3    & 1275.1   & 2.1\%       & 7.5         & $\geq6.7\times10^{16}$ \\
 Resonant $2L$             & $0\nu$     & $2^+$ 1431.0    & 1319.8   & 2.1\%       & 6.1         & $\geq8.2\times10^{16}$ \\

 ~                         & ~           & ~             & ~           & ~          & ~         & ~                     \\
 $^{192}$Os$~\to^{192}$Pt  & ~           & ~             & ~           & ~          & ~         & ~                     \\
 $2\beta^{-}$              & $2\nu+0\nu$ & $2^+$ 316.5 & 316.5       & 3.0\%      & 27        & $\geq5.3\times10^{19}$  \\

 \hline
\end{tabular}
}
\end{center}
\end{table*}

\section{CONCLUSIONS}

The radiopurity of an osmium sample has been investigated and
the first experiment to search for $2\beta$ processes in
$^{184}$Os and $^{192}$Os was carried out by using ultra-low
background HPGe $\gamma$ spectrometry. After 2741 h of data taking
with a 173 g ultra-pure osmium sample limits on double-beta
processes in $^{184}$Os have been established at the level of
$T_{1/2}\sim (10^{14}-10^{17}$) yr. A possible resonant neutrinoless
double-electron capture in $^{184}$Os to the excited 1322.2 keV,
1386.3 keV and 1431.0 keV states of $^{184}$W are bound at the
level of $\lim T_{1/2}\sim 10^{16}$ yr. The $2\beta^{-}$ decay of
$^{192}$Os to the first excited level of $^{192}$Pt is restricted
to $T_{1/2}\geq 5.3\times10^{19}$ yr at 90\% CL.

The osmium sample has subsequently been placed in a well-type ultra
low background HPGe detector especially designed for low
energy $\gamma$ rays spectrometry\footnote{The main goal of this stage
of the experiment is to search for $\alpha$ decay of $^{184}$Os and $^{186}$Os
to the excited levels of their daughter nuclei (see \cite{Bell12} where an indication
for $\alpha$ decay of $^{184}$Os to the first excited level (103.5 keV) of $^{180}$W
with the half-life $T_{1/2}\sim5\times10^{14}$ yr was obtained).}. In such a way
we expect to
improve the sensitivity of the experiment at least to the $2\nu2K$
decay, which is the most probable double-beta process in
$^{184}$Os. Further progress can be achieved by using osmium
enriched in $^{184}$Os, although this will require new enrichment techniques to be developed.

\section{ACKNOWLEDGMENTS}

The authors from the Institute for Nuclear Research (Kyiv,
Ukraine) were supported in part by the Space Research Program of
the National Academy of Sciences of Ukraine.


\begin{thebibliography}{99}
\bibitem{Zdes02}   Yu.G.~Zdesenko, Rev. Mod. Phys. 74 (2002) 663.
\bibitem{Elli02}   S.R.~Elliot, P.~Vogel, Ann. Rev. Nucl. Part. Sci. 52 (2002) 115.
\bibitem{Verg02}   J.D.~Vergados, Phys. Rep. 361 (2002) 1.
\bibitem{Elli04}   S.R.~Elliot, J.~Engel, J. Phys. G 30 (2004) R183.
\bibitem{Avig05}   F.T.~Avignone III, G.S.~King, Yu.G.~Zdesenko, New J. Phys. 7 (2005) 6.
\bibitem{Ejir05}   H.~Ejiri, J. Phys. Soc. Japan 74 (2005) 2101.
\bibitem{Avig08}   F.T.~Avignone III, S.R.~Elliott, J.~Engel, Rev. Mod. Phys. 80 (2008) 481.
\bibitem{Klap08}   H.V.~Klapdor-Kleingrothaus, Int. J. Mod. Phys. E 17 (2008) 505.
\bibitem{Rode11}   W.~Rodejohann, Int. J. Mod. Phys. E 20 (2011) 1833.
\bibitem{Elli12}   S.R.~Elliott, Mod. Phys. Lett. A 27 (2012) 1230009.
\bibitem{Verg12}   J.D.~Vergados, H.~Ejiri, F.~Simkovic, Rep. Prog. Phys. 75 (2012) 106301.
\bibitem{Hirs94}   M.~Hirsch et al., Z. Phys. A 151 (1994) 347.
\bibitem{Smor12}   C.~Smorra et al., Phys. Rev. C 86 (2012) 044604.
\bibitem{Berg11}   M.~Berglund, M.E.~Wieser, Pure Appl. Chem. 83 (2011) 397.
\bibitem{Audi03}   G.~Audi, A.H.~Wapstra, C.~Thibault, Nucl. Phys. A 729 (2003) 337.
\bibitem{Kriv11}   M.I.~Krivoruchenko et al., Nucl. Phys. A 859 (2011) 140.
\bibitem{Bagl10a}  C.M.~Baglin, Nucl. Data Sheets 111 (2010) 275.
\bibitem{Bagl10b}  C.M.~Baglin, Nucl. Data Sheets 113 (2012) 1871.
\bibitem{Bell12}   P.~Belli et al., First search for double beta decay of osmium by low background HPGe detector, to be published in Proc. of 4$^{th}$ Int. Conf. NPAE-Kyiv2012, 3--7 September 2012, Kyiv, Ukraine.
\bibitem{Azha00}   V.M. Azhazha, G.P. Kovtun, G.F. Tihinsky, The obtaining and metallophysics of high-pure metals, Metallofizika i Noveishie Tekhnologii 22 (2000) 21 (in Russian).
\bibitem{Lide03}   D.R.~Lide (ed), CRC Handbook of Chemistry and Physics, 84th Edition. CRC Press. Boca Raton, Florida, 2003.
\bibitem{EGSnrc}   I. Kawrakow and D.W.O. Rogers, NRCC report PIRS-701 (2003)
\bibitem{Agos03}   S.~Agostinelli et al., Nucl. Instrum. Meth. A 506 (2003) 250;
\bibitem{Alli06}   J.~Allison et al., IEEE Trans. Nucl. Sci. 53 (2006) 270.
\bibitem{DECAY0}   O.A.~Ponkratenko et al., Phys. At. Nucl. 63 (2000) 1282; \\
                   V.I.~Tretyak, to be published.
\bibitem{ToI98}    R.B.~Firestone et al., \textit{Table of Isotopes}, 8-th ed., John Wiley, New York,

                   1996 and CD update, 1998.
\bibitem{Feld98}   G.J.~Feldman, R.D.~Cousins, Phys. Rev. D 57 (1998) 3873.
\bibitem{Kibe08}   T.~Kib$\acute{{\rm e}}$di et al., Nucl. Instrum. Meth. A 589 (2008) 202; \\
                   http://www.nndc.bnl.gov/nndcscr/ensdf$_-$pgm/analysis/BrIcc/.
\bibitem{Dre02}    M. Drescher et al., Nature 419 (2002) 803.
\bibitem{Wint55}     R.G. Winter, Phys. Rev. 100 (1955) 142.


\end{thebibliography}
\end{document}